\documentstyle[preprint,prl,aps]{revtex}
\begin{document}
\tightenlines
\draft

\title{Mixed states in a sparsely encoded associative memory model storing ultrametric patterns}
\author{Tomoyuki Kimoto}
\address{
Oita National College of Technology,
1666,Maki,Oita-shi,Oita 870-0152,Japan}

\author{Masato OKADA}
\address{
Brain Science Institute, RIKEN, 2-1 Hirosawa, Wako-shi, Saitama 351-0198, Japan,\\
ERATO Kawato Dynamic Brain Project, 2-2 Hikaridai, Seika-cho, Soraku-gun, Kyoto 619-0288, Japan
}

\date{\today}
\maketitle

%
%-------------------Abstract---------------------------------------------------
%
\begin{abstract}
When mixed states are composed of $s$ memory patterns, 
$s$ types of mixed states, which can become equilibrium states of the model, can be generated. 
We previously reported that
storage capacities for most mixed states 
composed of {\it uncorrelated} memory patterns do not diverge
at the sparse limit of the firing rate, $f \rightarrow 0$.
On the contrary to the uncorrelated case,
we show that the storage capacities for all mixed states 
composed of {\it correlated} memory patterns diverge as $1/|f \log f|$ 
even when the correlation of memory patterns is infinitestimal small.
We also show that,
when the firing rate is fixed,
as the correlation coefficient increases, 
the storage capacities of the mixed states composed of all correlated memory patterns increase, 
while those of the mixed states composed of only some of the correlated memory patterns decrease.
\end{abstract}

\section{Introduction}
When an associative memory model stores memory patterns 
as a result of correlation learning, the states generated 
by the mixing of arbitrary memory patterns as well as 
the stored memory patterns automatically become 
equilibrium states of the model. 
These states are called {\it mixed states}, 
and they are not simply a side effect that is unnecessary for information processing. 
For example, Amari's proposed ``concept formation model'' 
uses the stability of mixed states (Amari 1977). 
The correlated attractor (Griniasty et al. 1993; Amit et al. 1994), 
which is a model of Miyashita's findings (Miyashita 1988a), 
is considered a mixed state in a broad sense. 
Moreover, Parga and Rolls and Elliffe et al. used mixed states 
in their research on the mechanism of invariant recognition 
with a coordinate transformation in the visual system (Parga and Rolls 1998; Elliffe et al. 1999).

In many of the studies on the mixed states of associative memory models, 
the stored memory patterns were not correlated.
However, 
if we use {\it ultrametric} memory patterns, which have hierarchical correlation, 
a relationship between the memory items encoded in the memory patterns is naturally implemented by the correlation coefficient between the memory patterns.
This is one of remarkable advantages for using a distributed representation in the associative memory model.
Amari proposed an associative memory model 
that stores ultrametric memory patterns (Amari 1977), 
and Fontanari analyzed the properties of this model with respect to the replica theory (Fontanari 1990). 
Applicability of the replica theory is limited,
since it cannot be applied to a system in which the free energy cannot be defined (e.g., a model with a nonmonotonic output function). 
Therefore, 
Toya and Okada focused on self-consistent signal-to-noise analysis (SCSNA) (Shiino and Fukai 1992), 
which can be used to analyze the properties of a model in which the free energy cannot be defined. 
They extended conventional SCSNA to generalized SCSNA 
so that it can be used to treat a model that stores ultrametric memory patterns and has a general class of neuron output function (Toya et al. 2000).
Moreover,
they analyzed the mixed states of an associative memory model 
storing ultrametric memory patterns using the generalized SCSNA.

In the studies mentioned above,
the memory patterns with a firing rate of $50\%$ was used. 
However the sparse coding scheme is thought to be used in the brain 
as indicated by certain physiological findings (Miyashita 1988b) 
and theoretical findings (Tsodyks and Feigel'man 1988; Buhmann et al. 1989; Amari 1989; Perez-Vicente and Amit 1989; Okada 1996).
Thus, it is necessary to analyze the properties of the mixed states in a sparsely encoded associative memory model. 
We previously analyzed the mixed states in a sparsely encoded associative memory model storing non-ultrametric memory patterns (Kimoto and Okada 2001), 
and found that when $s$ memory patterns are mixed, 
$s$ types of mixed states can be generated and that all of them can be made 
an equilibrium state by adjusting the threshold value of the model.
Moreover, among the $s$ types, 
the OR mixed state generated by OR operation on each element of the memory patterns performs best at the sparse limit of the firing rate, $f \rightarrow 0$.
This is because at the sparse limit, 
the storage capacity of the OR mixed state diverges as $1/|f \log f|$, 
and the storage capacities of the other mixed states become 0.

We have now added a sparse coding scheme to an associative memory model 
storing the ultrametric memory patterns and analyzed the properties of the model using a statistical mechanical method. 
We found that the storage capacities of all mixed states diverge as $1/|f \log f|$ at the sparse limit.

%-----------------------------------------------------------------------------
\section{Model}
We consider an associative memory model composed of $N$ neurons with output function $\Theta(\cdot)$. 
We use synchronous dynamics:
\begin{eqnarray}
 x_i^{t+1} &=& \Theta(\sum_{j \ne i}^N J_{ij} x_j^t + h^t),
                \quad \quad i=1,2 \cdots,N, \\
                \label{eq-model1:equilibrium}
  \Theta(u) &=& \cases{
		1 & u $\ge$ 0 \cr
        	0 & u  $<$  0 \cr
      		} ,
	\label{eq-model1:theta}
\end{eqnarray}
where $x_i^t$ represents the state of the $i$th neuron at discrete time $t$, 
and $J_{ij}$ denotes the synaptic coupling from the $j$th neuron to the $i$th neuron. 
The threshold value, $h^t$, is assumed to be independent of serial number $i$ of a neuron. 
Its concrete value is described later. 
The output function, $\Theta(\cdot)$, is assumed to be a step function, as shown in Eq. (\ref{eq-model1:theta}).

Memory patterns $\mbox{\boldmath $\eta$}^{\mu\nu}$ 
are the ultrametric memory patterns with two-step hierarchy ($\mu$ and $\nu$),
where $\mu$ is the number of group to which the memory pattern belongs, 
and $\nu$ is the number of memory pattern in that group.
Each group includes the $s$ memory patterns generated with correlation, 
while the memory patterns of different groups are generated without correlation.
Many procedures can be used to generate the set of ultrametric memory patterns;
we use the following procedure. 
First, parent pattern $\mbox{\boldmath $\eta$}^\mu$, the parent of the memory patterns of the group $\mu$,
 is generated. 
the parent pattern $\mbox{\boldmath $\eta$}^\mu$ is a vector of $N$ dimensions composed of the elements 0 and 1, 
and each component $\eta^{\mu}_i$ is independently generated using firing rate $f$:
\begin{eqnarray}
	&& \mbox{Prob}[\eta^\mu_i=1] = 1-\mbox{Prob}[\eta^\mu_i=0] = f,
	\nonumber \\
	&& i=1,2,\cdots,N.
\label{eq-model1:pattern1}
\end{eqnarray}
Next, 
the memory patterns, $\mbox{\boldmath $\eta$}^{\mu \nu}$ $(1 \le \nu \le s)$, 
are generated based on the parent pattern $\mbox{\boldmath $\eta$}^\mu$.
Each component $\eta^{\mu \nu}_i$ is determined based on $\eta^{\mu}_i$ with probabilities $K$ and $R$:
\begin{eqnarray}
	&& \mbox{Prob}[\eta^{\mu \nu}_i=1] 
          = 1-\mbox{Prob}[\eta^{\mu \nu}_i=0] 
          = \cases{
	      K & :$\eta^\mu_i=1$ \cr
              R & :$\eta^\mu_i=0$ \cr
            }
	\nonumber \\
	&& i=1,2,\cdots,N, \nu=1,2,\cdots,s.
\label{eq-model1:pattern2}
\end{eqnarray}
Many groups are generated with the repetition of this procedure. 
If the probabilities $K$ and $R$ satisfy the equation 
\begin{equation}
     R=f \frac{1-K}{1-f},
\label{eq-model1:P_rev}
\end{equation}
the firing rate of the memory patterns $\mbox{\boldmath $\eta$}^{\mu \nu}$ becomes $f$.
To set the firing rate of the memory patterns $\mbox{\boldmath $\eta$}^{\mu \nu}$ to $f$,
we use Eq. (\ref{eq-model1:P_rev}).
The memory patterns belonging to the same group are mutually correlated,
since they are generated from the same parent pattern. 
The memory patterns belonging to different groups are mutually uncorrelated,
since they are generated from different parent patterns.

We calculate the correlation coefficient between the memory patterns thus generated.
Since the mean value, $\mbox{E}[\eta^{\mu \nu}_i]$, of the memory pattern is $f$,
the correlation coefficient between the memory patterns is the term on the left side of Eq. (\ref{eq-model1:pattern_similarity}). 
By calculating the correlation coefficient at the limit of $N \rightarrow \infty$ 
using Eqs. (\ref{eq-model1:pattern1}), (\ref{eq-model1:pattern2}), and (\ref{eq-model1:P_rev}), 
the coefficient is equivalent to the term on the right side of Eq. (\ref{eq-model1:pattern_similarity}).
\begin{eqnarray}
\frac{\mbox{E}[(\eta^{\mu\nu}_i-f)(\eta^{\mu'\nu'}_i-f)]}
	{\sqrt{\mbox{E}[(\eta^{\mu\nu}_i-f)^2]}
	 \sqrt{\mbox{E}[(\eta^{\mu'\nu'}_i-f)^2]}}
	&=& \cases{
	      1 & : $\mu=\mu', \nu=\nu'$  \cr
	      a & : $\mu=\mu', \nu \ne \nu'$ \cr
              0 & : $\mu \ne \mu'$ \cr
            }
	\label{eq-model1:pattern_similarity} \\
	a &=& (\frac{K-f}{1-f})^2
	\label{eq-model1:a}
\end{eqnarray}
The correlation matrix of the ultrametric memory patterns is as shown in Fig.1. 
The correlation coefficient becomes $a$ in same group and becomes 0 in different groups. 
According to Eq. (\ref{eq-model1:a}), 
the memory patterns belonging to the same group are exactly same $(a=1)$ when $K=1$, 
and they become uncorrelation $(a=0)$ when $K=f$.

The synaptic coupling is determined by the following learning method:
\begin{equation}
	J_{ij}=\frac {1}{Nf(1-f)} 
	\sum_{\mu=1}^{\alpha N} \sum_{\nu=1}^{s}
	(\eta_i^{\mu\nu} - f) (\eta_j^{\mu\nu} - f),
	\label{eq-model1:covariance}
\end{equation}
where $\alpha N$ is the number of stored groups and $\alpha$ is the loading rate. 

Next, we explain the ``mixed state''. 
We consider the mixed state composed of the $s$ memory patterns belonging to the same group. 
We explain generation method of a mixed state by using, as an example, 
the memory patterns $\mbox{\boldmath $\eta$}^{1, 1},\mbox{\boldmath $\eta$}^{1, 2},\cdots, \mbox{\boldmath $\eta$}^{1,s}$ belonging to the first group. 
The $i$th element of the mixed state is set to $'1'$ if the number of firing state $'1'$ is $k$ or more in the $i$th elements $\eta^{1,1}_i,\eta^{1,2}_i,\cdots,\eta^{1,s}_i$.
Otherwise, it is set to $'0'$.
There are thus $s$ types of mixed states because $1 \leq k \leq s$. 
We call this mixed state $\mbox{\boldmath{$\gamma$}}^{(s,k)}$. 
Among the $s$ types of mixed states, 
mixed state $\mbox{\boldmath{$\gamma$}}^{(s,1)}$ is considered to be an OR mixed state;
its $i$th element is given by OR operation on the $i$th elements of the $s$ memory patterns.
Mixed state $\mbox{\boldmath{$\gamma$}}^{(s,s)}$ is considered to be an AND mixed state; 
its $i$th element is given by AND operation on the $i$th elements.
Mixed state $\mbox{\boldmath{$\gamma$}} ^{(s,[\frac{s+1}{2}])} $ is a majority decision mixed state;
its $i$th element is given by majority of the $i$th element values 0 and 1 of the $s$ memory patterns.
($[\cdot]$ stands for the Gauss symbol.)
These mixed states can be made to automatically become equilibrium states 
by setting an appropriate threshold value, 
even when they are not learned.

The threshold value, $h$, in Eq. (\ref{eq-model1:equilibrium}) is 
calculated using the firing rate of the recalled pattern. 
The threshold value obtained by this method corresponds approximately to the optimum threshold value at which the storage capacity is maximized (Okada 1996).  
Since the firing rate of the equilibrium state is $f$ when the memory pattern is recalled, the threshold value is determined by solving
\begin{equation} 
 f = \frac{1}{N} \sum_{i=1}^N
  \Theta \left(\sum_{j \neq i}^N J_{ij} x_j^t + h^t \right).
\label{eq-model1:threshold}
\end{equation}
When recalling the mixed state $\mbox{\boldmath{$\gamma$}} ^{(s,k)}$, $f$ in Eq. (\ref{eq-model1:threshold}) is replaced with firing rate $f^{(s,k)}$ of the mixed state $\mbox{\boldmath{$\gamma$}} ^{(s,k)}$:
\begin{eqnarray} 
 f^{(s,k)} &=& \mbox{E}[\gamma^{(s,k)}_i] \nonumber \\
           &=& \sum_{n=k}^{s} {}_s \mbox{C}_{n} 
    [f K^{n} (1-K)^{s-n} + (1-f) R^{n} (1-R)^{s-n}],
  \label{eq-model1:f_s_k}
\end{eqnarray}
where $\mbox{C}$ is the number of combinations.

The overlap between the equilibrium state $\mbox{\boldmath{$x$}}$ 
and the memory pattern $\mbox{\boldmath{$\eta$}}^{\mu \nu}$ is defined as
\begin{equation}
  m^{\mu\nu} = \frac{1}{Nf(1-f)}
  \sum_{i=1}^N (\eta_i^{\mu\nu}-f) x_i.
  \label{eq-model1:overlap}
\end{equation}
If the equilibrium state $\mbox{{\boldmath{$x$}}}$ is exactly equal to $\mbox{\boldmath{$\eta$}}^{\mu \nu}$, $m_\mu=1$. 
The overlap between the equilibrium state {\boldmath{$x$}} and the mixed state $\mbox{\boldmath{$\gamma$}} ^{(s,k)}$ is defined in a similar manner:
\begin{equation}
  M^{(s,k)} = \frac{1}{N f^{(s,k)} (1- f^{(s,k)} )}
  \sum_{i=1}^N (\gamma^{(s,k)}_i - f^{(s,k)}) x_i.
  \label {eq-model1:overlapM}
\end{equation}
If the equilibrium state $\mbox{{\boldmath{$x$}}}$ is exactly equal to $\mbox{\boldmath{$\gamma$}} ^{(s,k)}$, $M^{(s,k)}=1$.

%------------------------------------------------------------------------------
\section{Results}
\subsection{Qualitative evaluation using 1step-S/N analysis}

Whether the mixed states become the equilibrium states of the model can easily be examined using the 1step-S/N analysis (Kimoto and Okada 2001). 
In this section, we examine the stability of the mixed state $\mbox{\boldmath $\gamma$}^{(s,k)}$ using the 1step-S/N analysis.

If an initial state $\mbox{\boldmath $x$}^0$ of the model is set to a mixed state $\mbox{\boldmath $\gamma$}^{(s,k)}$, composed of the fist group memory patterns $\mbox{\boldmath $\eta$}^{1,1},\mbox{\boldmath $\eta$}^{1,2},\cdots,\mbox{\boldmath $\eta$}^{1,s}$, the internal potential $u_i$ of the $i$th neuron can be described using Eqs. (\ref{eq-model1:covariance}) and (\ref{eq-model1:overlap}):
\begin{eqnarray}
 u_i &=& \sum_{j \neq i}^N J_{ij} \gamma^{(s,k)}_j + h \nonumber \\
 &=& \sum_{\nu=1}^{s} (\eta^{1,\nu}_i-f) m^{1,\nu} +h +\bar{z_i},\\
 m^{1,\nu} &=& \frac{1}{Nf(1-f)} \sum_{i=1}^N
  (\eta^{1,\nu}_i -f) \gamma^{(s,k)}_i, 
	\label{eq.S/N2_m} \\
 \bar{z_i} &=& \frac{1}{Nf(1-f)} \sum_{\mu=2}^{\alpha N} \sum_{\nu=1}^{s} 
      \sum_{j \ne i}^{N} (\eta^{\mu \nu}_i-f)(\eta^{\mu \nu}_j-f) \gamma^{(s,k)}_j .
      \label{eq.S/N2_noise}
\end{eqnarray}
At the limit $N \rightarrow \infty$, the overlaps become the same, $m^{1,1}=m^{1,2}=,\cdots,=m^{1,s}$, because the memory patterns, $\mbox{\boldmath $\eta$}^{1,1},\mbox{\boldmath $\eta$}^{1,2},\cdots,\mbox{\boldmath $\eta$}^{1,s}$, are at the same distance from the mixed state $\mbox{\boldmath $\gamma$}^{(s,k)}$. 
By placing these overlaps with $m^{(s,k)}$, $u_i$ becomes  
\begin{equation}
   u_i = m^{(s,k)} \sum_{\nu=1}^{s} (\eta^{1,\nu}_i-f) +h +\bar{z_i}. 
\label{eq.S/N2}
\end{equation}
The first term is a signal term to recall the mixed state $\mbox{\boldmath $\gamma$}^{(s,k)}$.
The second term is the threshold value, and the third term is cross-talk noise, which prevents recall of the mixed state.
At the limit $N \rightarrow \infty$, $\bar{z_i}$ obeys the Gaussian distribution $N(0,\alpha f^{(s,k)} s)$ because $\eta_i^{\mu\nu},\eta_j^{\mu\nu},$ and $\gamma_j^{(s,k)}$ in Eq. (\ref{eq.S/N2_noise}) are mutually independent. 
Here, $f^{(s,k)}$ is the firing rate of the mixed state; it is given by Eq. (\ref{eq-model1:f_s_k}).
Note that the first term in Eq. (\ref{eq.S/N2}) is the linear sum of the memory patterns belonging to the first group.
Therefore, the state of neuron $\Theta(u_i)$ can be made the mixed state $\mbox{\boldmath $\gamma$}^{(s,k)}$ 
when $m^{(s,k)}>0$, $h$ is set appropriately, and the variance of cross-talk noise is small sufficiently.

We next discuss the stability of the mixed states at the sparse limit, $f \rightarrow 0$.
First, we evaluate the variance of cross-talk noise, $\alpha f^{(s,k)} s$.
When the firing rate of the memory pattern becomes $f \rightarrow 0$, 
that of the mixed state becomes $f^{(s,k)} \rightarrow 0$ and the variance of cross-talk noise becomes $\alpha f^{(s,k)} s \rightarrow 0$.
Since crosstalk noise, which prevents recall of the mixed state, 
disappears at $f \rightarrow 0$, we need only evaluate $m^{(s,k)}>0$ to estimate the stability of the mixed state.
Therefore, we next evaluate the value of $m^{(s,k)}$.
First, we rewrite $m^{(s,k)}$ in Eq. (\ref{eq.S/N2_m}) as a function of $f,s,k,K,$ and $R$
using Eqs. (\ref{eq-model1:pattern1}) and (\ref{eq-model1:pattern2}):
\begin{eqnarray}
  m^{(s,k)}=\frac{1}{f(1-f)} 
\Big[ 
 \!\!\!\!\!\! && \sum_{i=k}^{s-1} {}_{s-1}\mbox{C}_i f (1-K)^{s-1-i} K^i (-f) \nonumber \\
 &+& \sum_{i=k-1}^{s-1} {}_{s-1}\mbox{C}_i f (1-K)^{s-1-i} K^{i+1} (1-f) \nonumber \\
&+& \sum_{i=k}^{s-1} {}_{s-1}\mbox{C}_i (1-f) (1-R)^{s-i} R^i (-f) \nonumber \\
&+& \sum_{i=k-1}^{s-1} {}_{s-1}\mbox{C}_i (1-f) (1-R)^{s-1-i} R^{i+1} (1-f) 
\Big].
 \label{m(sk)_1}
\end{eqnarray}
Next,
we derive $m^{(s,k)}$ at $f \rightarrow 0$.
By substituting $R=f\frac{1-K} {1-f}$ in Eq. (\ref{eq-model1:P_rev}) for Eq. (\ref{m(sk)_1}) 
and setting $f \rightarrow 0$, the first term, the third term, and part of the fourth term can be omitted:
\begin{eqnarray}
  m^{(s,k)} &=& \sum_{i=k-1}^{s-1} {}_{s-1}\mbox{C}_i (1-K)^{s-1-i} K^{i+1} \nonumber \\
&+& \sum_{i=k-1}^{s-1} {}_{s-1}\mbox{C}_i (1-K) (0)^i .
 \label{m(sk)_2}
\end{eqnarray}
The second term in Eq. (\ref{m(sk)_2}) becomes 1 only at $i=0$ ($k=1$) and becomes 0 at $i \ge 1$ ($k \ge 2$).
The following equations are obtained by deriving $m^{(s,k)}$ separately for $k=1$ and $k \ge 2$. 
\begin{eqnarray}
  k=1 &:& m^{(s,k)} = \sum_{i=k-1}^{s-1} {}_{s-1}\mbox{C}_i (1-K)^{s-1-i} K^{i+1} +(1-K) =1\\
  k \ge 2&:& m^{(s,k)} = \sum_{i=k-1}^{s-1} {}_{s-1}\mbox{C}_i (1-K)^{s-1-i} K^{i+1}
 \label{m(sk)_3}
\end{eqnarray}
We rewrite Eq. (\ref{m(sk)_3}) as follows using the correlation coefficient between memory patterns,
($a=(\frac{K-f}{1-f})^2 \rightarrow K^2$ at $f \rightarrow 0$):
\begin{eqnarray}
  k=1&:& m^{(s,k)} = 1, \label{m(sk)_4a} \\
  k \ge 2&:& m^{(s,k)} = \sum_{i=k-1}^{s-1} {}_{s-1}\mbox{C}_i (1-\sqrt{a})^{s-1-i} (\sqrt{a})^{i+1}.
 \label{m(sk)_4b}
\end{eqnarray}
The $k=1$ means the overlap for the OR mixed state, and the $k \ge 2$ means the overlap for the other mixed states.
Therefore, $m^{(s,k)}$ for the OR mixed state becomes 1 at $f \rightarrow 0$, and $m^{(s,k)}$ for the other mixed states depends on $a$ .
Let's consider the value of overlap $m^{(s,k)}$ in Eq. (\ref{m(sk)_4b}).
Though $m^{(s,k)}=0$ at $a=0$, when $a$ has a non-zero value, even a small one,
$m^{(s,k)}$ takes a non-zero value.
When the memory patterns have nocorrelation ($a=0$), 
the storage capacities of the mixed states, 
except for the OR mixed state, become 0 
because the signal term becomes 0 at the sparse limit, $f \rightarrow 0$ 
(Kimoto and Okada 2001).
However,
when the memory patterns are even slightly correlated ($a>0$), the storage capacities for all mixed states may diverge because the signal term becomes non-zero at the sparse limit, $f \rightarrow 0$.

We have discussed the qualitative mechanism of the stability of mixed states 
when the memory patterns are correlated. 
In the next section, we quantitatively analyze the storage capacities for each type of mixed state.

%------------------------------------------------------------------------------
\subsection{Quantitative evaluation using generalized SCSNA}

Here 
we quantitatively analyze the storage capacities of the mixed states using the generalized SCSNA
(Shiino and Fukai 1992, Toya et al. 2000). 
The order parameter equations of the equilibrium state derived using the SCSNA correspond to those of the equilibrium state derived using the replica theory of the statistical mechanics.

We consider the case in which the equilibrium state $\mbox{\boldmath $x$}$ has non-zero overlaps, $m^{1, \nu} = \frac{1}{Nf(1-f)} \sum_{i=1}^{N} (\eta^{1, \nu}_i-f) x_i$, with $s$ memory patterns, $\mbox{\boldmath $\eta$}^{1,\nu}  (1 \le \nu \le s)$.
This means that the memory pattern belonging to the first group or the mixed state generated by the memory patterns of the first group is recalled. 
To derive the SCSNA order parameter equations of the present model, 
we first transform the synaptic coupling $J_{ij}$ into the following form, which can easily be applied to the SCSNA.
That is, 
$J_{ij}$ is divided into a term related to the memory patterns of the first group and another term.
\begin{eqnarray}
  J_{ij} &=& \frac {1}{Nf(1-f)} \sum_{\nu=1}^{s}
	      (\eta^{1,\nu}_i-f) (\eta^{1,\nu}_j-f)      \nonumber \\
          & & + \frac {1}{Nf(1-f)} \sum_{\mu=2}^{\alpha N} \sum_{\nu=1}^{s}
	      (\eta_i^{\mu \nu} - f) (\eta_j^{\mu \nu} - f)
\label{eq-model1:covariance2}
\end{eqnarray}
We next introduce a set of i.i.d. patterns, $\sigma^{\mu \nu}_i$, to the $(\eta^{\mu \nu}_i-f)$:
\begin{eqnarray}
   \mbox{E}[\sigma^{\mu \nu}_i] &=& 0, \\
   \mbox{E}[\sigma^{\mu \nu}_i \sigma^{\mu \nu'}_i] &=& 0, \quad (\nu \ne \nu').
\label{eq-model1:E_sigma}
\end{eqnarray}
The second term of the synaptic coupling in Eq. (\ref{eq-model1:covariance2}) is rewritten using $\sigma^{\mu \nu}_i$ as
\begin{eqnarray}
   J_{ij} &=& \frac {1}{Nf(1-f)} \sum_{\nu=1}^{s}
	      (\eta^{1,\nu}_i-f) (\eta^{1,\nu}_j-f)      \nonumber \\
          & & + \frac {1}{Nf(1-f)} \sum_{\mu=2}^{\alpha N}
              \sum_{\nu=1}^{s} \sum_{\nu'=1}^{s}
	      \sigma^{\mu \nu}_i A_{\nu \nu'} \sigma^{\mu \nu'}_j,
\label{eq-model1:covariance3}
\end{eqnarray}
where
$A_{\nu \nu'}$ is the correlation coefficient 
between the memory patterns $\mbox{\boldmath{$\eta$}}^{\mu \nu}$ and $\mbox{\boldmath{$\eta$}}^{\mu \nu'}$ of the same group.

Let ${\bf e}^\nu$ $(\nu=1,2,\cdots,s)$ be a set of normalized eigenvectors 
of the $s \times s$ dimensional matrix ${\mbox{\bf A}}$ composed of $A_{\nu \nu'}$.
We introduce a set of rotated patterns, $\bar{\mbox{\boldmath{$\sigma$}}}^\mu_i=(\bar{\sigma}^{\mu,1}_i,\bar{\sigma}^{\mu,2}_i,\cdots,\bar{\sigma}^{\mu,s}_i)$:
\begin{eqnarray}
  	\mbox{\boldmath{$\sigma$}}^\mu_i &=& {\bf T} \bar{\mbox{\boldmath{$\sigma$}}}^\mu_i,
  \label{eq-model1:rotated_pattern}\\
	{\bf T} &=& ({\bf e}^1,{\bf e}^2,\cdots,{\bf e}^{s}).
  \label{eq-model1:eigenvectors}
\end{eqnarray}
Since the rotated patterns $\bar{\mbox{\boldmath{$\sigma$}}}^\mu_i$ 
are simply rotated using the matrix ${\mbox{\bf T}}$, 
the distribution is the same as that of $\mbox{\boldmath{$\sigma$}}^\mu_i$.
Using $\bar{\mbox{\boldmath{$\sigma$}}}^\mu_i$, 
we rewrite the synaptic coupling $J_{ij}$ in Eq. (\ref{eq-model1:covariance3}) as
\begin{eqnarray}
   J_{ij} &=& \frac {1}{Nf(1-f)} \sum_{\nu=1}^{s} 
	      (\eta^{1,\nu}_i-f) (\eta^{1,\nu}_j-f)      \nonumber \\
          & & + \frac {1}{Nf(1-f)} \sum_{\mu=2}^{\alpha N}
              \sum_{\nu=1}^{s} 
	      \lambda^\nu \bar{\sigma}^{\mu \nu}_i \bar{\sigma}^{\mu \nu}_j,
     \label{eq-model1:covariance4}
\end{eqnarray}
where
$\lambda^\nu$ is the eigenvalue of matrix ${\mbox{\bf A}}$ for normalized eigenvector ${\bf e}^\nu$:
$\lambda^1=1+(s-1)(\frac{K-f}{1-f})^2$C$\lambda^\nu=1-(\frac{K-f}{1-f})^2,(2 \le \nu \le s)$.
Internal potential $u_i$ is written as follows, using $J_{ij}$ from Eq. (\ref{eq-model1:covariance4}) 
and the overlap $m^{1,\nu}$ from Eq. (\ref{eq-model1:overlap}):
\begin{eqnarray}
   u_i &=& \sum_{j \ne i}^N J_{ij} x_j \nonumber \\
       &=& \sum_{\nu=1}^{s} (\eta^{1,\nu}_i-f) m^{1,\nu}
           +\sum_{\mu=2}^{\alpha N} \sum_{\nu=1}^{s} 
	      \lambda^\nu \bar{\sigma}^{\mu \nu}_i \bar{m}^{\mu \nu}
           - \alpha s x_i,
\end{eqnarray}
where
\begin{equation}
    \bar{m}^{\mu \nu} = \frac{1}{Nf(1-f)} 
          \sum_{i=1}^{N} \bar{\sigma}^{\mu \nu}_i x_i.
\end{equation}
Using the SCSNA, we can easily derive the SCSNA order parameter equations.
\begin{eqnarray}
  Y_i
   &=& \Theta \left(\sum_{\nu=1}^{s} (\eta^{1,\nu}_i-f) m^{1,\nu}
	  + \Gamma Y_i + \sqrt{\alpha r} z +h^t
	  \right) \label{eq-model1:Y2} \\
 m^{1,\nu} &=& 
	\frac{
  \int_{-\infty}^{\infty} D_z <(\eta^{1,\nu}_i-f) Y_i>_{\mbox{\boldmath $\eta$}}
	}
	{f(1-f)}
   \label{eq-model1:m2} \\
 q &=& \int_{-\infty}^{\infty} D_z <(Y_i)^2>
  _{\mbox{\boldmath $\eta$}} \label {eq-model1:q2} \\
 U &=& \frac{1}{\sqrt{\alpha r}}
  \int_{-\infty}^{\infty} D_z z <Y_i>_{\mbox{\boldmath $\eta$}}
  \label {eq-model1:U2}
  \\
 D_z &=& \frac{dz}{\sqrt{2 \pi}} \exp(-\frac{z^2}{2})
  \label {eq-model1:D_z2}
  \\
 r &=& q \sum_{\nu=1}^{s} \frac{(\lambda^\nu)^2}{(1-\lambda^\nu U)^2}
  \label {eq-model1:r2} \\
 \Gamma &=& \alpha \sum_{\nu=1}^{s} 
     \frac{ (\lambda^\nu)^2 U}{1-\lambda^\nu U}
  \label{eq-model1:Gamma2}
\end{eqnarray}
The $< \cdots >_{\mbox{\boldmath $\eta$}}$ stands for an ensemble average over the first group memory patterns,
$\mbox{\boldmath $\eta$}=({\bf \eta}_i^{1,1}, {\bf \eta}_i^{1,2}, \cdots, {\bf \eta}_i^{1,s})$.
Equation (\ref{eq-model1:Y2}) may have more than one solution according to Maxwell's ``equal area rule'', 
which was originally applied to thermodynamics.
According to this rule, Eq. (\ref{eq-model1:Y2}) can be rewritten as
\begin{equation}
  Y_i = \Theta \left(\sum_{\nu=1}^{s} (\eta^{1,\nu}_i-f) m^{1,\nu}
	  + \frac{1}{2}\Gamma  + \sqrt{\alpha r} z +h^t
	  \right).
	\label{eq-model1:Y3} \\
\end{equation}
We can obtain the following SCSNA order parameter equations 
by integrating Eqs. (\ref{eq-model1:m2})-(\ref{eq-model1:Y3}).
\begin{eqnarray}
  m^{1,\nu} &=&	< 
		\frac{\eta^{1,\nu}-f}{2f(1-f)} 
	  	\mbox{erf}(
                      \frac{ \sum_{\nu=1}^s (\eta^{1,\nu}-f) m^{1,\nu}
                      +h^t +\frac{\Gamma}{2}}
		{\sqrt{2 \alpha r}}
	)>_{\mbox{\boldmath $\eta$}}, \nonumber \\
    && \qquad\qquad\qquad\qquad\qquad\qquad\qquad\qquad 1 \le \nu \le s 
 \label{eq-model1:m3} \\
 q &=& \frac{1}{2} + \frac{1}{2} <
		\mbox{erf}(\frac{
                    \sum_{\nu=1}^s  (\eta^{1,\nu}-f) m^{1,\nu}
                    +h^t +\frac{\Gamma}{2}}
		{\sqrt{2 \alpha r}}
	>_{\mbox{\boldmath $\eta$}} 
 \label{eq-model1:q3} \\
 U &=& \frac{1}{\sqrt{2\pi \alpha r}}< 
		\mbox{exp}(-(\frac{
                       \sum_{\nu=1}^s (\eta^{1,\nu}-f) m^{1,\nu}
                       +h^t +\frac{\Gamma}{2}}
		{\sqrt{2 \alpha r}})^2
	>_{\mbox{\boldmath $\eta$}}
 \label{eq-model1:U3} \\
 r &=& q \sum_{\nu=1}^{s} \frac{(\lambda^\nu)^2}{(1-\lambda^\nu U)^2}
  \label {eq-model1:r3} \\
 \Gamma &=& \alpha \sum_{\nu=1}^{s} \frac{(\lambda^\nu)^2 U}{1-\lambda^\nu U}
  \label{eq-model1:Gamma3}
\end{eqnarray}
We can then derive the relationship between $m^{1,\nu}$ and $\alpha$ by solving simultaneous equations (\ref{eq-model1:m3})-(\ref{eq-model1:Gamma3}).
Note that the threshold value $h$ must be determined first because some of the equations include $h$. 
The following order parameter equation, which determines the threshold value,
is derived from Eq. (\ref{eq-model1:threshold}).
\begin{eqnarray}
    f &=& \int_{-\infty}^{\infty} D_z <Y_i>_{\mbox{\boldmath $\eta$}} \nonumber \\
      &=& \frac{1}{2} + \frac{1}{2} <
		\mbox{erf}(\frac{
                    \sum_{\nu=1}^s (\eta^{1,\nu}-f) m^{1,\nu}
                    +h^t +\frac{\Gamma}{2}} {\sqrt{2 \alpha r}}
	>_{\mbox{\boldmath $\eta$}} 
        \label {eq-model1:threshold2}
\end{eqnarray}
To recall the mixed state $\mbox{\boldmath{$\gamma$}} ^{(s,k)}$,
the threshold value is determined using the following order parameter equation, 
in which $f$ in Eq. (\ref{eq-model1:threshold2}) is replaced with the firing rate, $f^{(s,k)}$, of the mixed state.
\begin{equation}
    f^{(s,k)} = \frac{1}{2} + \frac{1}{2} <
		\mbox{erf}(\frac{
                    \sum_{\nu=1}^s (\eta^{1,\nu}-f) m^{1,\nu}
                    +h^t +\frac{\Gamma}{2}} {\sqrt{2 \alpha r}}
	>_{\mbox{\boldmath $\eta$}} 
        \label {eq-model1:threshold3}
\end{equation}
The overlap between the equilibrium state $\mbox{\boldmath $x$}$ 
and the mixed state $\mbox{\boldmath $\gamma$}^{(s,k)}$ 
is derived from Eq. (\ref{eq-model1:overlapM}):
\begin{equation}
  M^{(s,k)} = 
	< \frac{\gamma^{(s,k)}-f^{(s,k)}} {2f^{(s,k)}(1-f^{(s,k)})}
  	  	\mbox{erf}(\frac{ 
                         \sum_{\nu=1}^s (\eta^{1,\nu}-f) m^{1,\nu}
			 +h^t +\frac{\Gamma}{2}}
		{\sqrt{2 \alpha r}}
	)>_{\mbox{\boldmath $\eta$}}. 
  \label {eq-model1:M}
\end{equation}

Next, 
to investigate the effectiveness of the generalized SCSNA when applied to the present model,
 we compared the results of the SCSNA with those of a computer simulation. 
Figure 2 shows the overlaps $m^{1,1}, m^{1,2},$ and $m^{1,3}$ 
for various loading rates $\alpha$ when recalling the memory pattern $\mbox{\boldmath $\eta$}^{1,1}$.
The data points and error bars show the results of the computer simulation, 
and the lines connecting the data points show the results of the SCSNA.
We used $f$=0.1, $a=0.25$, and $s=3$.
In the computer simulation, the number of neurons, $N$, was set to 10,000, 
and the simulation was run 11 times for each parameter.
The data points show the median values, and the ends of the error bars show the $1/4$ 
and $3/4$ deviations.
The results show that the overlap $m^{1,1}$ is nearly equal to $1$ at $\alpha \simeq 0$, 
so the memory pattern $\mbox{\boldmath $\eta$}^{1,1}$ was recalled; 
$m^{1,2}$ and $m^{1,3}$ are 0.25, 
since the correlation coefficient between $\mbox{\boldmath $\eta$}^{1,1}$ and 
$\mbox{\boldmath $\eta$}^{1,2},\mbox{\boldmath $\eta$}^{1,3}$ is $0.25$.
The results of the SCSNA show that the overlap $m^{1,1}$ gradually decreases from 1 
as the loading rate is increased from 0, 
and that the equilibrium state disappears at $\alpha \simeq 0.078$. 
The storage capacity is thus $\alpha_c \simeq 0.078$.
The results of the computer simulation correspond to those of the SCSNA 
for $0 \le \alpha \le 0.078$; the equilibrium state became unstable for $\alpha > 0.078$.
The results of the SCSNA and computer simulation also correspond well 
for the overlaps $m^{1,2}$ and $m^{1,3}$.

Figure 3 shows the overlap $M^{(3,1)}$ for various loading rates  
when recalling the OR mixed state $\mbox{\boldmath $\gamma$}^{(3,1)}$.
We again used $f=0.1$, $a=0.25$, and $s=3$.
The data points and error bars again show the results of the computer simulation, 
and the lines connecting the data points show the results of the SCSNA.
The overlap $M^{(3,1)}$ is nearly equal to $1$ at $\alpha \simeq 0$, 
so the OR mixed state $\mbox{\boldmath $\gamma$}^{(3,1)}$ is recalled. 
The equilibrium state disappears at $\alpha \simeq 0.036$, 
so the storage capacity is $\alpha_c \simeq 0.036$.
Since the results of the SCSNA are correspond fairly well with those of the computer simulation, 
as shown in Figs. 2 and 3, 
we used the SCSNA to examine the properties of the memory pattern and of the mixed states. 

Figure 4 shows the storage capacities of the memory pattern 
and of the mixed states for various values of $a$ for $s=3$ and $f$=0.01.
Since $a=0$ is equivalent to a conventional model storing non-ultrametric memory patterns, 
the storage capacity is equal to that of the conventional model.
At $a=0$, the storage capacities of the memory pattern and of the OR mixed state
are larger than those of the majority decision mixed state and the AND mixed state.
As $a$ increases, 
though, the storage capacities of the majority decision mixed state and the AND mixed state rapidly increase, 
more than the storage capacities of the memory pattern and the OR mixed state.

To analyze the storage capacities of the memory pattern 
and the mixed states at the sparse limit, $f \rightarrow 0$, 
we plotted the asymptotes of the storage capacities for $a=0,$ $0.001,$ and $0.25$ (Fig. 5).
Since the present model corresponds to the conventional model at $a=0$, 
the asymptotes of the memory pattern and of the OR mixed state diverge as $1/|f \log f|$ at the sparse limit, 
and those of the majority decision mixed state and of the AND mixed state approach 0, 
as we previously reported (Kimoto and Okada 2001).
At $a=0.001$ and $a=0.25$, 
the asymptotes of the memory pattern and of all mixed states diverge as $1/|f \log f|$ at the sparse limit.
Thus, 
when the memory patterns are even slightly correlated, 
the storage capacities for all of the mixed states diverge at the sparse limit.

Figure 6 shows the threshold values $h_c$ for various values of $a$ 
when the loading rate $\alpha$ reached the storage capacity $\alpha_c$.
The threshold values were calculated using Eqs. (\ref{eq-model1:threshold2}) and (\ref{eq-model1:threshold3}) with $f=0.01$ and $s=3$.
The figure shows that the mixed state, which becomes the equilibrium state of the model,
can be changed by adjusting the threshold value.

When the model of $s=3$ becomes that of $s=4$ due to the addition of a correlated memory pattern,
how does the storage capacity of a mixed state composed of three memory patterns change?
To answer this question, we examined the storage capacity of a mixed state composed of all four memory patterns belonging to the same group, that of a mixed state  composed of three of the four memory patterns belonging to the same group, and that of a mixed state composed of two of the four memory patterns belonging to the same group, all in the $s=4$ model.
Figure 7 shows the results for OR mixed states. 
In the OR mixed state composed of all four memory patterns belonging to the same group, $\alpha_c \rightarrow 4.2$ at $a \rightarrow 1$.
In the OR mixed state composed of only some of the memory patterns belonging to the same group, $\alpha_c \rightarrow 0$ at $a \rightarrow 1$.
The same properties were obtained for the other mixed states (results not shown).
Thus, if a memory pattern is added to a group, the mixed state composed of all memory patterns, including the added memory pattern, becomes stable instead of the mixed state that had previously been stable.

%------------------------------------------------------------------------------
\section{Conclusion}

We have examined a sparsely encoded associative memory model 
storing ultrametric memory patterns which have hierarchical correlation.
We previously reported that
storage capacities for most mixed states composed of uncorrelated memory patterns do not diverge
at the sparse limit of the firing rate, $f \rightarrow 0$, in our paper (Kimoto and Okada 2001).
On the contrary to the uncorrelated case,
we have now found that the storage capacities for all mixed states 
composed of correlated memory patterns diverge as $1/|f \log f|$ 
even when the correlation of memory patterns is infinitestimal small.
We have also found that 
as the correlation coefficient increases, the storage capacity of a mixed state composed of all memory patterns belonging to same group increases, while that of mixed states composed of only some of the memory patterns belonging to a same group decreases.
Furthermore we have found that the mixed state, which becomes the equilibrium state of the model, can be changed by adjusting the threshold value.

%--------------------------------------------------------------
\noindent 
\section*{Acknowledgements}
This work was partially supported by a Grant-in-Aid 
for the scientific reserch basic(c) No. 14580438
and for the Encouragement of Young Scientists No. 14780309
from the Ministry of Education, Culture, Sports, Science 
and Technology of Japan. 

%%------------------------------------------------------------------------------

\end{document}

%%-----------Figure captions---------------------------------------------------

\begin{figure}[tb]
 \begin{center}
 \includegraphics[width=7cm,height=7cm]{fig1.eps}\\ 
 \end{center}
   \vspace{-7mm}
 \caption{Correlation matrix of ultrametric memory patterns.
The number of memory patterns belonging to the same group is $s=3$, 
the correlation coefficient between the memory patterns belonging to the same group is $a$, 
and the correlation coefficient between the memory patterns belonging to different groups is $0$.
} 
 \label{fig:1}
\end{figure}

\begin{figure}[tb]
 \begin{center}
 \includegraphics[width=7cm,height=7cm]{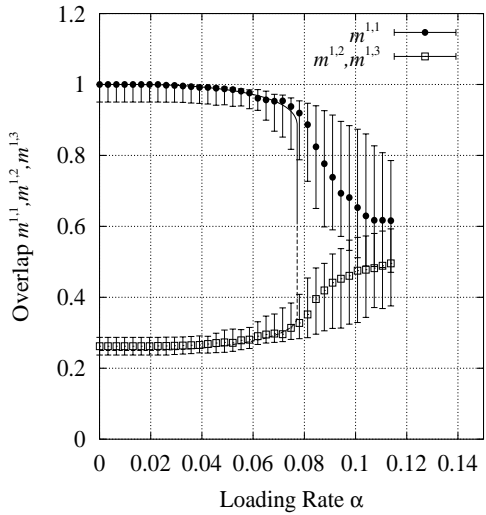}\\ 
 \end{center}
   \vspace{-7mm}
 \caption{
Overlap $m^{1,1}, m^{1,2}, m^{1,3}$ for various loading rates $\alpha$ when memory pattern $\mbox{\boldmath $\eta$}^{1,1}$ is recalled in $f=0.1, a=0.25, s=3$.
The solid line shows the result of the SCSNA for $m^{1,1}$ 
and the dashed line shows  the result of the SCSNA for $m^{1,2}, m^{1,3}$.
Data points and error bars show computer simulation results.
} 
 \label{fig:2}
\end{figure}

\begin{figure}[tb]
 \begin{center}
 \includegraphics[width=8cm,height=7cm]{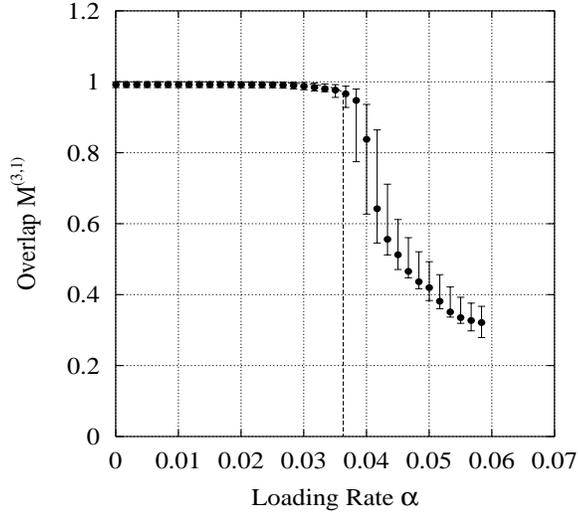}\\ 
 \end{center}
   \vspace{-7mm}
 \caption{Overlap $M^{(3,1)}$ for various loading rates $\alpha$ when OR mixed state $\mbox{\boldmath{$\gamma$}}^{(3,1)}$ is recalled in $f=0.1, a=0.25, s=3$.
The dashed line shows the result of the SCSNA.
Data points and error bars show computer simulation results.
} 
 \label{fig:3}
\end{figure}

\begin{figure}[tb]
 \begin{center}
 \includegraphics[width=7.5cm,height=7cm]{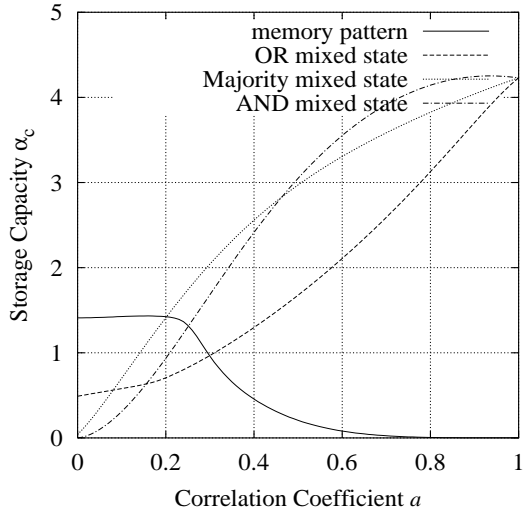}\\ 
 \end{center}
   \vspace{-7mm}
 \caption{
Storage capacity $\alpha_c$ for various values of correlation coefficient $a$ when $f=0.01, s=3$.
The storage capacities of the mixed states depending on the value of $a$.
} 
 \label{fig:4}
\end{figure}

\begin{figure}[tb]
 \begin{center}
 \includegraphics[width=8cm,height=7cm]{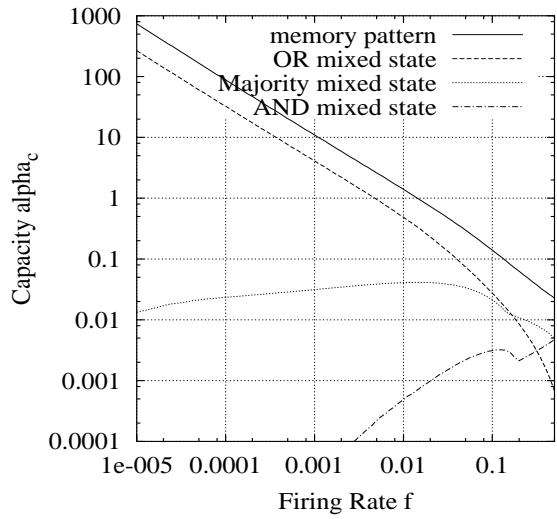} \\ (a) $a=0.0$ \\
 \includegraphics[width=8cm,height=7cm]{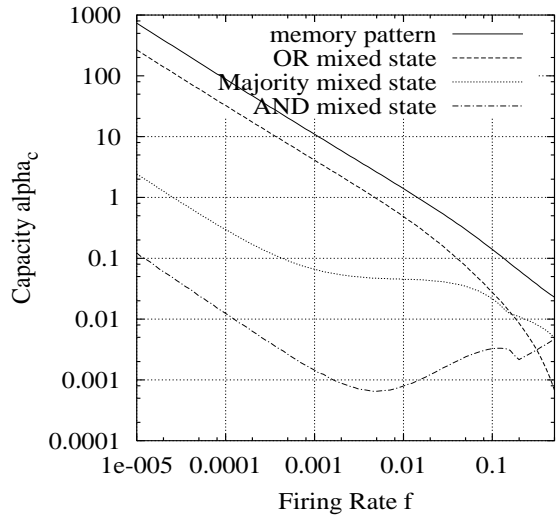} \\ (b) $a=0.001$ \\
 \includegraphics[width=8cm,height=7cm]{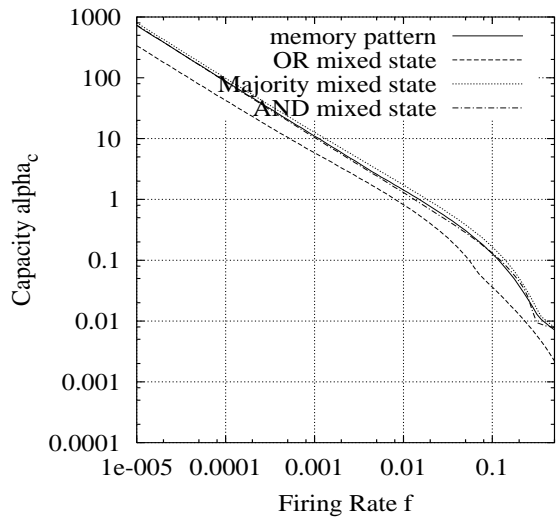} \\ (c) $a=0.25$ \\
 \end{center}
   \vspace{-7mm}
 \caption{
Asymptotic properties of storage capacity $\alpha_c$ when $a=0.0,0.001,0.25$ and $s=3$.
When the memory patterns are even slightly correlated ($a > 0$), the asymptotic properties of all mixed states diverge at $f \rightarrow 0$.
} 
 \label{fig:5}
\end{figure}

\begin{figure}[tb]
 \begin{center}
 \includegraphics[width=8cm,height=7cm]{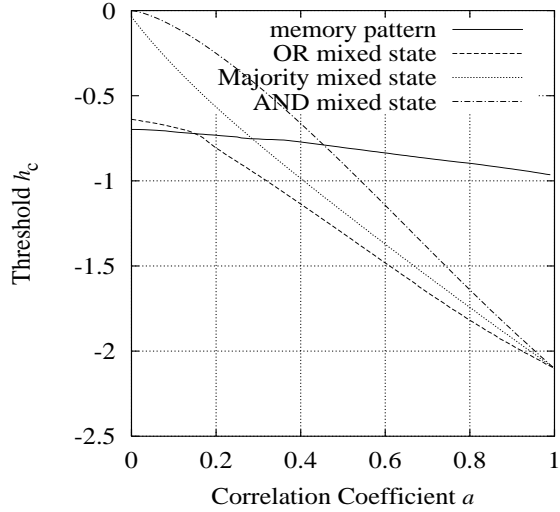}\\ 
 \end{center}
   \vspace{-7mm}
 \caption{
Threshold value $h_c$ for various values of correlation coefficient $a$ when $f=0.01, s=3$.
The mixed state, which becomes the equilibrium state of the model,
can be changed by adjusting the threshold value.
} 
 \label{fig:6}
\end{figure}

\begin{figure}[tb]
 \begin{center}
 \includegraphics[width=7cm,height=7cm]{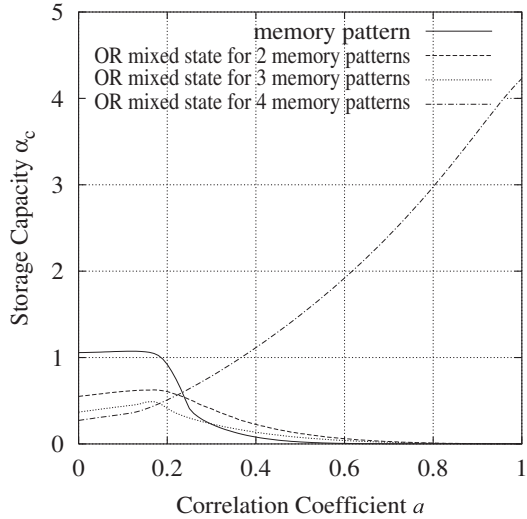}\\ 
 \end{center}
   \vspace{-7mm}
 \caption{
Storage capacity $\alpha_c$ of the OR mixed state for various values of correlation coefficient $a$ when $f=0.01, s=4$.
The storage capacity of the OR mixed state composed of 4(all) memory patterns belonging to same group become large as $a \rightarrow 1$.
However, the storage capacities of the other OR mixed states converges to 0 as $a \rightarrow 1$.
The same properties has been found also for the other mixed states (results not shown). 
} 
 \label{fig:7}
\end{figure}